\documentclass[nofootinbib,unsortedaddress]{revtex4-2}
\usepackage{amsmath}
\usepackage{dcolumn}
\usepackage{bm}
\usepackage{xcolor}
\usepackage{url}
\usepackage{graphicx}
\usepackage{float}
\usepackage{comment}

\usepackage[letterpaper,left=2.0cm,right=2.0cm,top=2.5cm,bottom=2.5cm]{geometry} 
\usepackage{siunitx} 
\usepackage{import} 
\usepackage{amssymb}
\usepackage{subcaption}

\usepackage[english]{babel}
\usepackage{hyperref}
\hypersetup{
    colorlinks=true,
    linkcolor=blue,
    filecolor=magenta,      
    urlcolor=blue,
    citecolor=blue,
}

\begin{document}

\title{Covariant interpretation of proper infall times in Kerr spacetime}

\author{Erick Past\'en}
\email{erick.contreras@usm.cl}
\affiliation{Departamento de F\'{\i}sica, Universidad de Santiago de Chile,
Avenida V\'{\i}ctor Jara 3493, Estaci\'on Central, 9170124, Santiago, Chile}

\author{Claudia \'Alvarez Rojas}
\email{claudia.alvarezr@usm.cl}
\affiliation{Departamento de F\'{\i}sica, Universidad T\'ecnica Federico Santa Mar\'{\i}a,
Av. Espa\~na 1680, Valpara\'{\i}so, Chile}

\author{Norman Cruz}
\email{norman.cruz@usach.cl}
\affiliation{Departamento de F\'{\i}sica, Universidad de Santiago de Chile,
Avenida V\'{\i}ctor Jara 3493, Estaci\'on Central, 9170124, Santiago, Chile}


\begin{abstract}
We investigate proper infall times in the Schwarzschild and Kerr spacetimes from a covariant perspective, focusing on the role of black--hole rotation in the focusing properties of timelike geodesic congruences. 
To perform a geometrically consistent comparison between rotating and non--rotating black holes, we analyse infall trajectories between surfaces of equal circumferential radius in the equatorial plane. 

Using equatorial timelike geodesics in the test--particle limit, we compute and compare the corresponding proper infall times for different values of the specific energy $E$, specific angular momentum $L$, and black--hole spin parameter $a$. 
Within the equal circumferential-radius prescription adopted here, we show that Kerr angular momentum $a$ can produce longer or shorter integrated proper infall times relative to the Schwarzschild case, depending on the orbital configuration and energy regime considered.

We then interpret these results within the covariant $1+3$ formalism of general relativity, in terms of the expansion, shear, and Raychaudhuri evolution of timelike congruences. 
Our analysis shows that the Kerr--Schwarzschild differences in proper infall times are encoded in the corresponding Raychaudhuri time integrand, which reflects a competition between the radial evolution of the expansion and the nonlinear focusing contribution driven by expansion and shear. 
Black--hole rotation modifies both effects in a systematic way, leading to distinct behaviours for prograde and retrograde infall configurations.
\end{abstract}


\maketitle

\section{Introduction}

In Newtonian gravity, angular momentum acts as a centrifugal barrier that opposes gravitational collapse and tends to increase infall times. Much of our physical intuition regarding gravitationally bound systems relies on this picture. In General Relativity, however, the interplay between energy, angular momentum, and spacetime curvature can substantially modify the dynamics of freely falling particles, particularly in strong-field regimes \citep{mtw1973gravitation,wald1984general,Carroll,Hartle}. Recent relativistic studies have emphasized that kinematics and motion can influence dynamical evolution in ways that differ qualitatively from Newtonian expectations. In cosmological contexts, for example, peculiar motions and relativistic energy fluxes can modify the growth and focusing properties of matter flows relative to the standard Newtonian picture \citep{Filippou_2021,TsapraziTsagas2020}. These results motivate the study of how relativistic rotation and motion affect characteristic dynamical timescales in strongly curved spacetimes. Rotating black-hole spacetimes therefore provide a natural setting in which to investigate how relativistic rotation influences infall dynamics and geodesic focusing.

The Schwarzschild and Kerr solutions constitute ideal laboratories for such investigations, since they allow one to isolate purely relativistic effects without the additional complications associated with hydrodynamics, radiation, or feedback processes \citep{schwarzschild,kerr1963,boyer-lind,carter,Chandrasekhar1983}. Previous investigations have shown that rotation can significantly modify the dynamics of infalling matter and relativistic accretion flows \citep{Das2007}, suggesting that the role of angular momentum in General Relativity is more subtle than its Newtonian counterpart.

At the same time, comparing distinct relativistic spacetimes is itself a nontrivial problem. Since different solutions of Einstein's equations do not admit a canonical point-by-point identification \citep{Geroch1969,Bengtsson2014}, physical comparisons necessarily require specifying which geometric or operational structures are held fixed. However, the absence of a unique prescription does not eliminate the physical relevance of such comparisons, but emphasizes the importance of defining geometrically meaningful observables when relating different spacetime geometries.

In this work, we study the free fall of test particles in the Schwarzschild and Kerr spacetimes from the perspective of integrated proper infall times and geodesic focusing. Motivated by the considerations above, we compare infall trajectories connecting surfaces of equal circumferential radius in the equatorial plane, providing a geometrically consistent prescription for relating Schwarzschild and Kerr motion.

Within this framework, we compute proper infall times for different values of the specific energy $E$, specific angular momentum $L$, and black-hole spin parameter $a$. Within the equal circumferential-radius comparison, we show that Kerr rotation can either increase or decrease the integrated proper infall time relative to Schwarzschild, depending on the orbital configuration and energy regime considered. The analysis is formulated within the covariant $1+3$ framework of General Relativity, where the Kerr--Schwarzschild differences in integrated proper infall times are interpreted through the Raychaudhuri evolution of timelike geodesic congruences. In particular, we show that the resulting infall dynamics are governed by the competition between the radial evolution of the expansion and the nonlinear focusing contribution associated with expansion and shear.

\section{Free--fall in Schwarzschild and Kerr black holes}\label{section:Metrics}

We work in geometric units $G=c=1$ and signature $(-,+,+,+)$. Timelike geodesics follow from the Lagrangian $\mathcal{L}=\tfrac12 g_{\mu\nu}\dot x^\mu \dot x^\nu$ with normalization $g_{\mu\nu}\dot x^\mu \dot x^\nu=-1$ (see e.g. \citep{wald1984general,Carroll,mtw1973gravitation}). Stationarity and axial symmetry imply two conserved quantities along the motion, the specific energy $E\equiv -p_t$ and azimuthal specific angular momentum $L\equiv p_\phi$, associated with the Killing vectors $\partial_t$ and $\partial_\phi$.

\subsection{Schwarzschild: effective potential and benchmark times}\label{subsec:Schw}

The Schwarzschild line element is \citep{schwarzschild}
\begin{equation}
ds^2=-\left(1-\frac{2M}{r}\right)dt^2+\left(1-\frac{2M}{r}\right)^{-1}dr^2+r^2(d\theta^2+\sin^2\theta\,d\phi^2).
\end{equation}
Without loss of generality we restrict to equatorial motion $\theta=\pi/2$. The conserved quantities can be written as
\begin{equation}
E=\left(1-\frac{2M}{r}\right)\dot t,\qquad L=r^2\dot\phi,
\end{equation}
and the radial equation takes the effective--potential form
\begin{equation}
\dot r^2 = E^2 - V_{\rm eff}^{\rm Schw}(r;L),\qquad
V_{\rm eff}^{\rm Schw}(r;L)=\left(1-\frac{2M}{r}\right)\left(1+\frac{L^2}{r^2}\right).
\label{eq:SchwRadial}
\end{equation}
Expanding the potential makes explicit the Newtonian limit and the leading relativistic correction,
\begin{equation}
V_{\rm eff}^{\rm Schw}(r;L)=1-\frac{2M}{r}+\frac{L^2}{r^2}-\frac{2ML^2}{r^3},
\end{equation}
Circular equatorial orbits which have zero radial velocity, satisfy $E^2=V_{\rm eff}$ and $dV_{\rm eff}/dr=0$, while stability is determined by $d^2V_{\rm eff}/dr^2$ (see e.g. \cite{wald1984general,Chandrasekhar1983}). 

For an ingoing trajectory ($\dot r<0$), the proper infall time from $r_0$ to $r_f$ is
\begin{equation}
\Delta\tau(r_0\to r_f)=\int_{r_f}^{r_0}\frac{dr}{\sqrt{E^2-V_{\rm eff}(r)}}.
\label{eq:TauGeneral}
\end{equation}
The corresponding coordinate time follows from $dt/d\tau=E/(1-2M/r)$, giving
\begin{equation}
\Delta t(r_0\to r_f)=\int_{r_f}^{r_0}\frac{E\,dr}{\left(1-\frac{2M}{r}\right)\sqrt{E^2-V_{\rm eff}(r)}}.
\label{eq:tcoordGeneral}
\end{equation}
In Schwarzschild coordinates $\Delta t$ diverges as $r_f\to 2M$, while $\Delta\tau$ remains finite, as is well known.

\begin{figure}[H]
    \centering
    \includegraphics[width=0.68\linewidth]{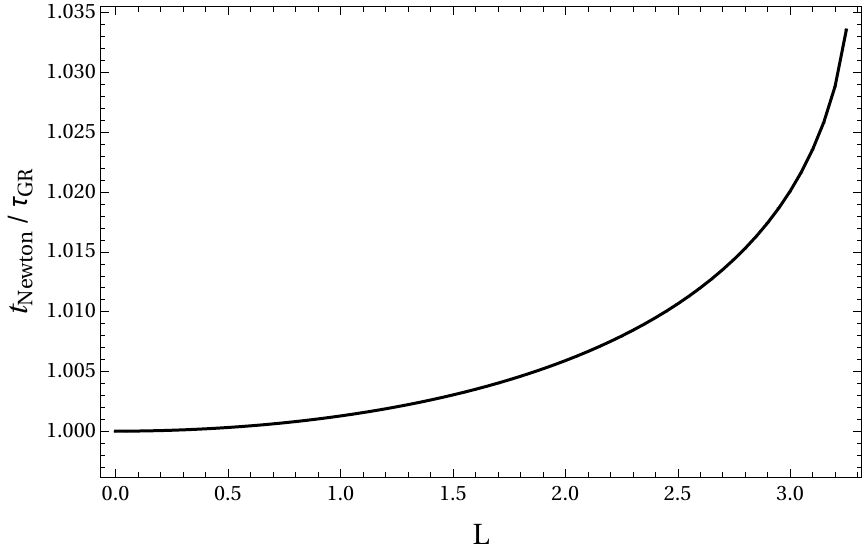}
    \caption{Ratio between the Newtonian infall time and the Schwarzschild proper infall time for trajectories from $r_0$ to $r_f=6M$, as a function of angular momentum $L$. The increasing deviation from unity quantifies the relativistic reduction of the integrated proper fall time relative to Newtonian expectations.}
    \label{fig:ratioN-GR}
\end{figure}

Before turning to Kerr, it is useful to establish a Schwarzschild baseline by comparing relativistic and Newtonian infall times for the same initial conditions. Figure~\ref{fig:ratioN-GR} shows the ratio $\Delta t_{\rm Newt}/\Delta\tau_{\rm GR}$ as a function of $L$ for infall from $r_0$ to $6M$ (which corresponds to the innermost stable circular orbit (ISCO) in Schwarzschild spacetime) for a particle starting with zero radial velocity.  As the particle angular momentum increases, the Newtonian infall time grows faster than the relativistic proper time. In other words, for fixed $(r_0,r_f)$ and comparable angular--momentum values, General Relativity predicts a shorter integrated infall time. This trend is made explicit in Fig.~\ref{fig:ratioN-GR}. 

\subsection{Kerr: equatorial geodesics and spin--dependent couplings}\label{subsec:Kerr}

The Kerr metric in Boyer--Lindquist coordinates is \citep{kerr1963,boyer-lind}
\begin{align}
ds^2 &= -\left(1-\frac{2Mr}{\Sigma}\right)dt^2-\frac{4Mar\sin^2\theta}{\Sigma}\,dt\,d\phi
+\frac{\Sigma}{\Delta}\,dr^2+\Sigma\,d\theta^2 \nonumber\\
&\quad+\left(r^2+a^2+\frac{2Ma^2r\sin^2\theta}{\Sigma}\right)\sin^2\theta\,d\phi^2,
\end{align}
with $\Sigma=r^2+a^2\cos^2\theta$ and $\Delta=r^2-2Mr+a^2$. We focus on equatorial motion $\theta=\pi/2$, for which $\Sigma=r^2$ and the Carter constant vanishes ($Q=0$; \cite{carter}). Conserved quantities are defined by
\begin{equation}
E=-p_t,\qquad L=p_\phi,\qquad p_\mu=g_{\mu\nu}\dot x^\nu.
\end{equation}
Solving the $(t,\phi)$ system yields $\dot t(E,L)$ and $\dot\phi(E,L)$ (e.g. \cite{Chandrasekhar1983,BardeenPressTeukolsky1972}), and the radial equation can be written in a compact form as
\begin{equation}
\dot{r}^{2} =
(E^{2} - 1)
+ \frac{2M}{r}
+ \frac{a^{2}(E^{2} - 1) - L^{2}}{r^{2}}
+ \frac{2M(aE - L)^{2}}{r^{3}},
\label{eq:KerrRadialExplicit}
\end{equation}
which reduces to the Schwarzschild expression when $a\to 0$. Equation~\eqref{eq:KerrRadialExplicit} shows explicitly how Kerr rotation modifies the radial geodesic dynamics relative to Schwarzschild through both spin-dependent quadratic terms and frame-dragging couplings.

The proper infall time in Kerr follows directly from the equatorial radial equation,
Eq.~\eqref{eq:KerrRadialExplicit}. For an ingoing trajectory ($\dot r<0$) we write
\begin{equation}
\frac{dr}{d\tau}=-\sqrt{\mathcal{R}(r;E,L,a)}\,,
\qquad
\mathcal{R}(r;E,L,a)\equiv
(E^{2} - 1)
+ \frac{2M}{r}
+ \frac{a^{2}(E^{2} - 1) - L^{2}}{r^{2}}
+ \frac{2M(aE - L)^{2}}{r^{3}},
\label{eq:KerrRdef}
\end{equation}
which can be equivalently cast in an effective--potential form $\dot r^{2}=E^{2}-V_{\rm}^{\rm Kerr}$, with
$V_{\rm}^{\rm Kerr}(r;E,L,a)\equiv E^{2}-\mathcal{R}(r;E,L,a)$.
Inverting Eq.~\eqref{eq:KerrRdef} gives an expression for the accumulated proper time between an initial radius
$r_0$ and a final radius $r_f<r_0$,
\begin{equation}
\Delta\tau(r_0\to r_f;E,L,a)
=
\int_{r_f}^{r_0}\frac{dr}{\sqrt{\mathcal{R}(r;E,L,a)}}\,,
\label{eq:TauKerrExplicit}
\end{equation}
which is finite for any $r_f$ outside the event-horizon and is evaluated numerically throughout this work. When $a\to 0$,
$\mathcal{R}$ reduces to the Schwarzschild expression and Eq.~\eqref{eq:TauKerrExplicit}, reproducing the curve
shown in Fig \ref{fig:ratioN-GR}.

\section{Integrated proper infall times}

Comparisons between different spacetime geometries require specifying the geometric structures that are kept fixed across the comparison, since no canonical point-by-point identification exists between distinct solutions of Einstein's equations \cite{Geroch1969,Bengtsson2014}. Motivated by this, we compare infall dynamics between Schwarzschild and Kerr spacetimes using trajectories connecting surfaces of equal circumferential radius in the equatorial plane.

We define the circumferential radius \cite{Hamilton2010} in Kerr spacetime through
\[
R_c^2 \equiv g_{\phi\phi},
\]
which in the equatorial plane becomes
\[
R_c(a)^2 = r^2 + a^2 + \frac{2Ma^2}{r}.
\]

For Schwarzschild spacetime, the circumferential radius reduces to the usual areal coordinate,
\[
R_c=r.
\]

Throughout this work, we compare trajectories between the same physical circumferential surfaces,
\[
R_0 = 10M,
\qquad
R_f = 3M.
\]
For each value of the Kerr spin parameter $a$, the corresponding Boyer--Lindquist radii $r_0(a)$ and $r_f(a)$ are obtained by solving the circumferential-radius relation above.

The integrated proper infall time is then computed through
\[
\Delta\tau(R_0\to R_f;E,L,a)
=
\int_{r_f(a)}^{r_0(a)}
\frac{dr}{\sqrt{\mathcal R(r;E,L,a)}},
\]
where $\mathcal R(r;E,L,a)$ is the radial geodesic potential associated with equatorial timelike motion in Kerr spacetime.

\begin{figure}[htbp]
  \centering
  \begin{subfigure}{0.49\textwidth}
    \centering
    \includegraphics[width=\textwidth]{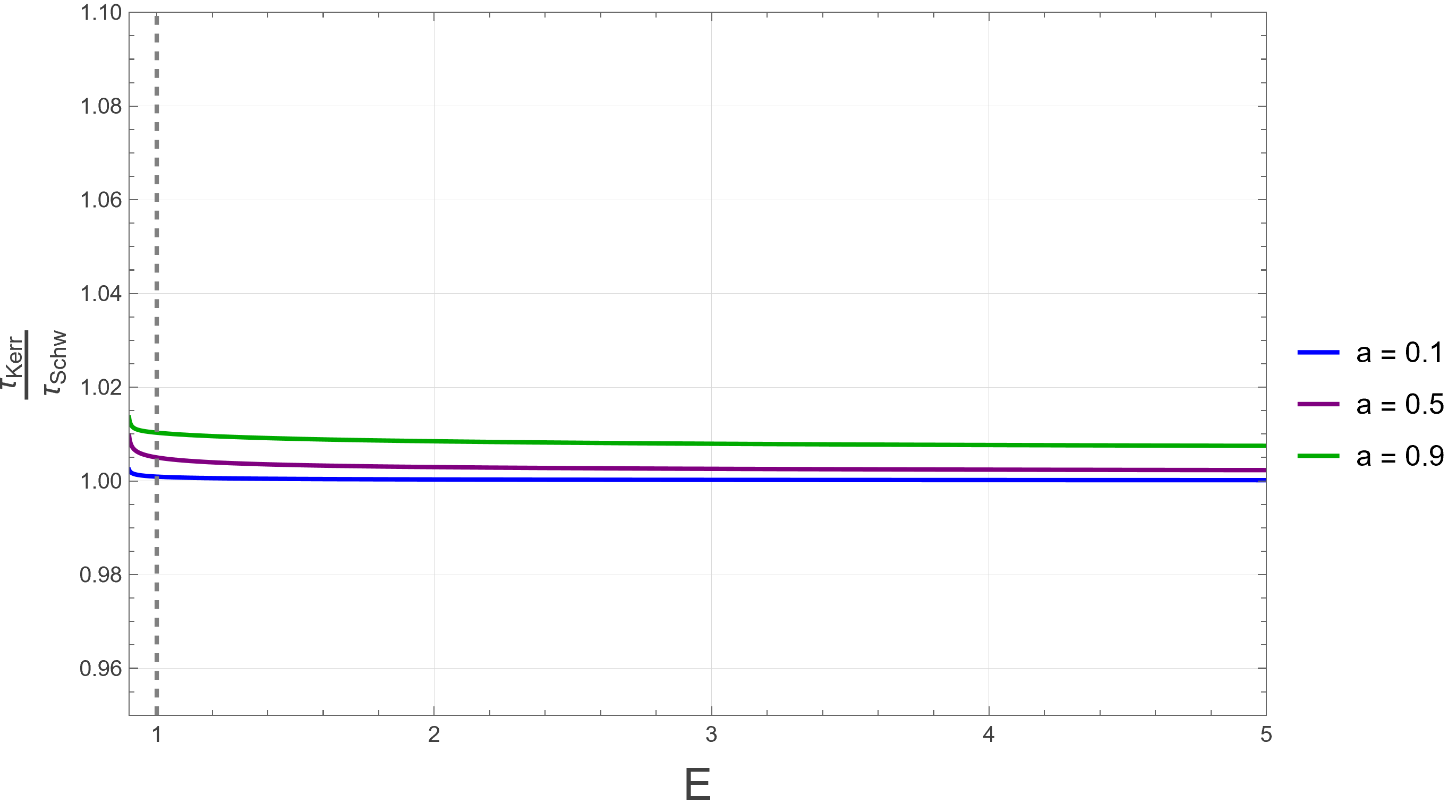}
    \caption{$L = 0.3$}
  \end{subfigure}
  \hfill
  \begin{subfigure}{0.49\textwidth}
    \centering
    \includegraphics[width=\textwidth]{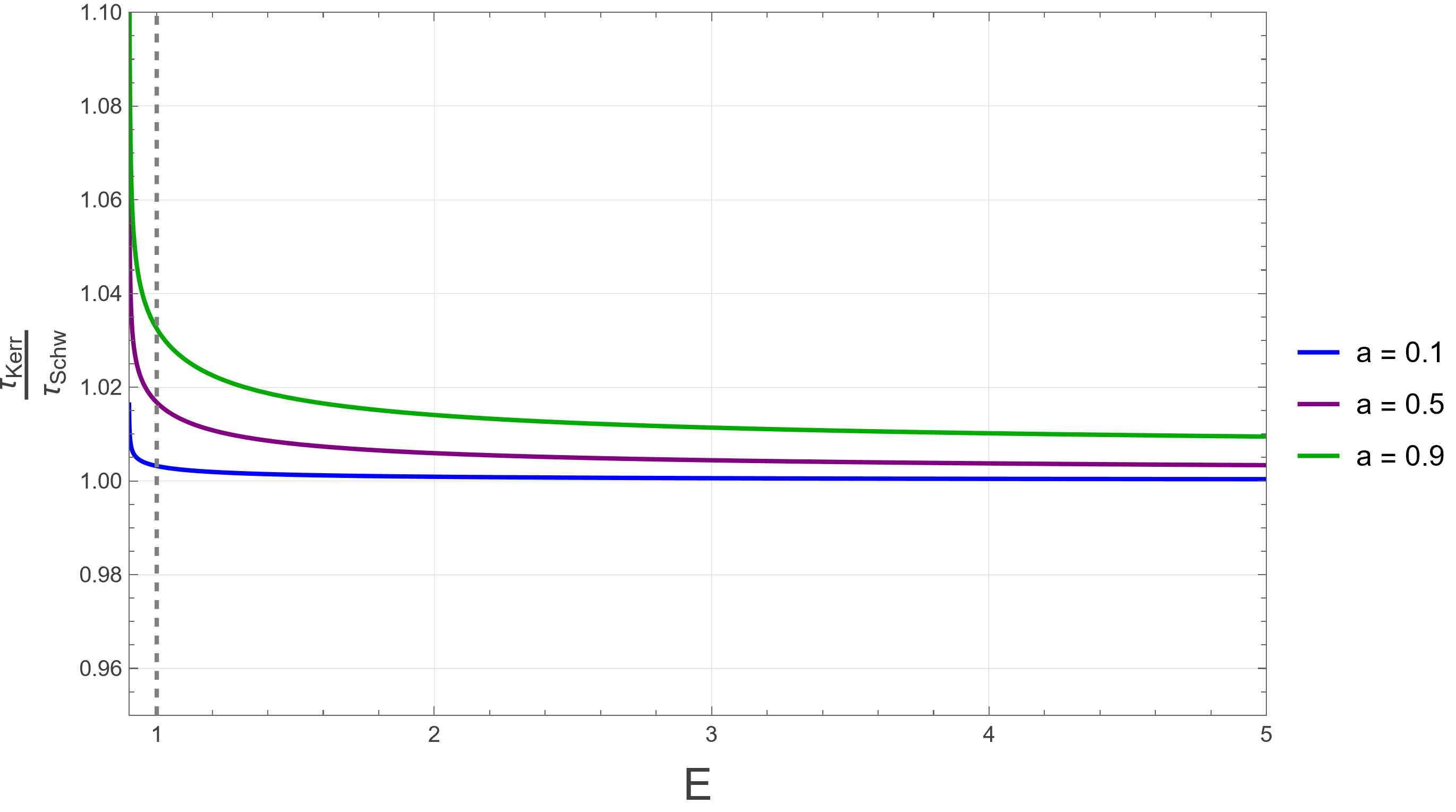}
    \caption{$L = 1$}
  \end{subfigure}
   \\[0.4em]
  \begin{subfigure}{0.49\textwidth}
    \centering
    \includegraphics[width=\textwidth]{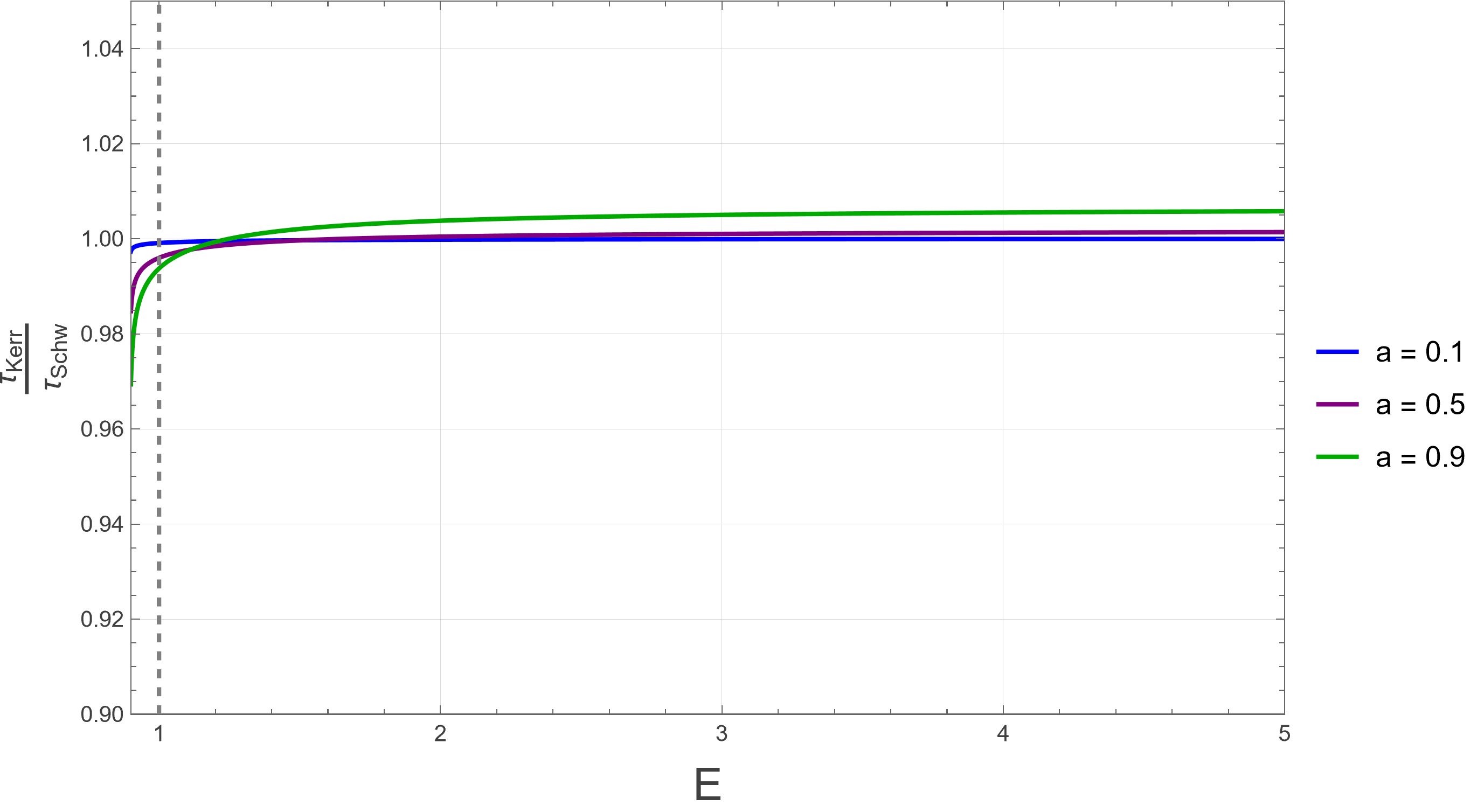}
    \caption{$L = -0.3$}
  \end{subfigure}
  \hfill
  \begin{subfigure}{0.49\textwidth}
    \centering
    \includegraphics[width=\textwidth]{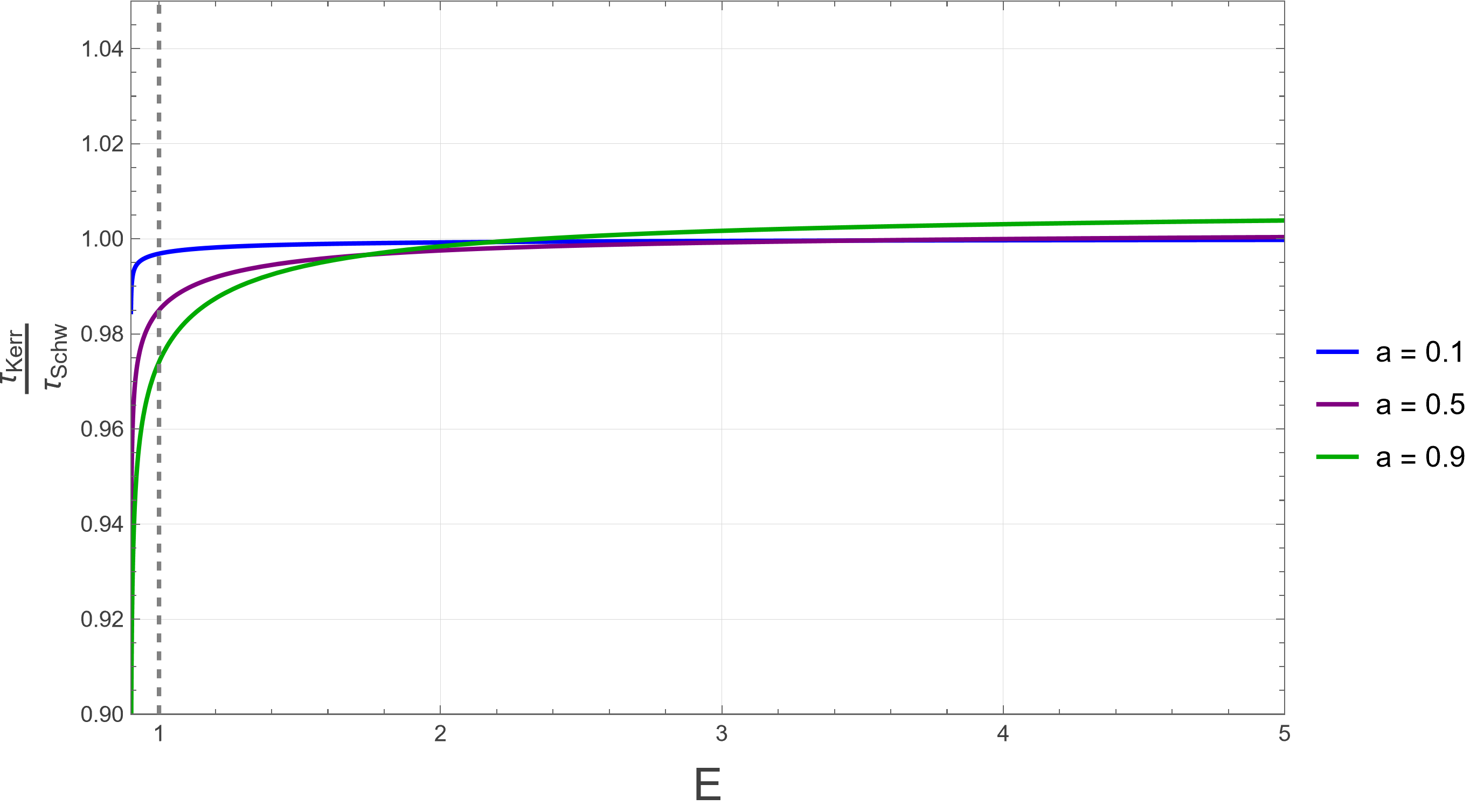}
    \caption{$L = -1$}
  \end{subfigure}
  \caption{
Ratio between Kerr and Schwarzschild integrated proper infall times for trajectories connecting equal circumferential surfaces $R_0=10M$ and $R_f=3M$ in the equatorial plane. The plots show the dependence of $\tau_{\rm Kerr}/\tau_{\rm Schw}$ on the specific energy $E$ for different values of the black--hole spin parameter $a$ and angular momentum $L$. Upper panels correspond to prograde configurations ($L>0$), while lower panels correspond to retrograde configurations ($L<0$). The results show that Kerr rotation modifies the integrated proper infall time depending on the orbital configuration and energy regime considered.
}
  \label{RATIO}
\end{figure}

Within the equal circumferential-radius prescription adopted here, figure~\ref{RATIO} shows that Kerr rotation can either increase or decrease the integrated proper infall time relative to Schwarzschild, depending on the orbital configuration and energy regime considered. For prograde trajectories ($L>0$), Kerr rotation generally increases the proper infall time relative to Schwarzschild, whereas retrograde trajectories ($L<0$) can exhibit shorter infall times at low energies before approaching or exceeding the Schwarzschild value at sufficiently large $E$. In all cases, the deviations become stronger as the black--hole spin parameter increases.

\section{Covariant 1+3 analysis }

A fully relativistic understanding of infall times requires going beyond coordinate velocities and effective potentials, and focusing instead on the covariant kinematics of geodesic congruences \citep{TSAGAS_2008,wald1984general,Poisson:2009pwt}. In General Relativity, the convergence or divergence of freely falling worldlines is encoded in the expansion scalar $\Theta$, whose evolution is governed by the Raychaudhuri equation \citep{Raychaudhuri1955}. This provides a direct connection between spacetime geometry, congruence focusing, and characteristic collapse times.

For a timelike geodesic congruence with four--velocity $u^a$, vanishing four--acceleration, and zero vorticity, the Raychaudhuri equation in vacuum spacetimes ($R_{ab}=0$) reduces to
\begin{equation}
\dot{\Theta}
=
-\frac{1}{3}\Theta^2
-
\sigma_{ab}\sigma^{ab},
\label{eq:raychaudhuri_simple}
\end{equation}
where $\sigma_{ab}$ is the shear tensor and the overdot denotes differentiation with respect to the proper time along the congruence. Throughout this work, we adopt the shear scalar
\begin{equation}
\sigma^2
\equiv
\frac12 \sigma^{ab}\sigma_{ab}.
\end{equation}

Equation~\eqref{eq:raychaudhuri_simple} shows that, in vacuum spacetimes, the local focusing of timelike congruences is entirely controlled by the competition between isotropic contraction, encoded in $\Theta$, and anisotropic distortion, encoded in $\sigma^2$.

Integrating the Raychaudhuri equation along the congruence (see e.g.~\citep{wald1984general,Poisson:2009pwt}) leads schematically to
\begin{equation}
\tau
\propto
\int
\frac{d\Theta}
{\Theta^2/3+2\sigma^2},
\label{tauWALD}
\end{equation}
which connects the integrated proper infall time to the cumulative focusing properties of the congruence.

Motivated by Eq.~\eqref{tauWALD}, we define the local Raychaudhuri time integrand
\begin{equation}
\mathcal I_\tau(R_c)
\equiv
\frac{d\Theta/dR_c}
{\Theta^2/3+2\sigma^2},
\label{eq:Itau}
\end{equation}
where $R_c$ is the circumferential radius used to compare Schwarzschild and Kerr trajectories. The numerator describes the radial evolution of the expansion, while the denominator measures the nonlinear focusing strength generated by the combined contribution of expansion and shear.

In terms of this quantity, the integrated proper infall time can be written schematically as
\begin{equation}
\tau
\propto
\int_{R_f}^{R_0}
\mathcal I_\tau(R_c)\, dR_c.
\label{eq:tau_integrand}
\end{equation}

\subsection{Congruence selection}

In the $1+3$ covariant formalism, a timelike congruence with four--velocity $u^a$
admits a family of spacelike hypersurfaces orthogonal to its worldlines if and only if
it satisfies the Frobenius integrability condition (see e.g.~\cite{ellis2012relativistic,wald1984general})
\begin{equation}
u_{[a}\nabla_b u_{c]} = 0 ,
\qquad \Longleftrightarrow \qquad
\omega_{ab} = 0 ,
\label{eq:frobenius}
\end{equation}
where the vorticity tensor is defined by
\begin{equation}
\omega_{ab} \equiv h_a{}^{c} h_b{}^{d}\,\nabla_{[d}u_{c]},
\qquad
h_{ab} \equiv g_{ab} + u_a u_b ,
\label{eq:vorticity_def}
\end{equation}
with $h_{ab}$ projecting orthogonally to $u^a$.

In stationary and axisymmetric spacetimes such as Schwarzschild and Kerr, the Killing vectors $\partial_t$ and $\partial_\phi$ imply two conserved quantities along geodesics: the specific energy $E\equiv -p_t$ and the specific angular momentum $L\equiv p_\phi$. We therefore consider congruences of freely falling test particles characterized by fixed values of $E$ and $L$.

For such congruences, the Frobenius condition \eqref{eq:frobenius} implies that fixing a uniform specific angular momentum does not necessarily introduce vorticity. Consequently, the congruences considered here can remain irrotational even though individual worldlines possess nonzero angular motion relative to asymptotic observers. In this case, rotational effects enter the Raychaudhuri dynamics through the shear and expansion of the congruence, rather than through intrinsic vorticity.

This property admits a simple Newtonian analogue: an axisymmetric flow with constant specific angular momentum,
\[
v_\phi = \frac{L}{\rho},
\]
has vanishing local vorticity away from the symmetry axis,
\[
(\nabla\times\mathbf v)_z
=
\frac{1}{\rho}
\frac{\partial}{\partial\rho}
(\rho v_\phi)
=0,
\]
despite the presence of nonzero angular motion. The relativistic condition $\omega_{ab}=0$ plays the direct analogue of this statement for the congruences studied here.

\subsection{Numerical results}

\begin{figure*}
    \centering
    \begin{subfigure}[t]{0.65\textwidth}
        \centering
        \includegraphics[width=\textwidth]{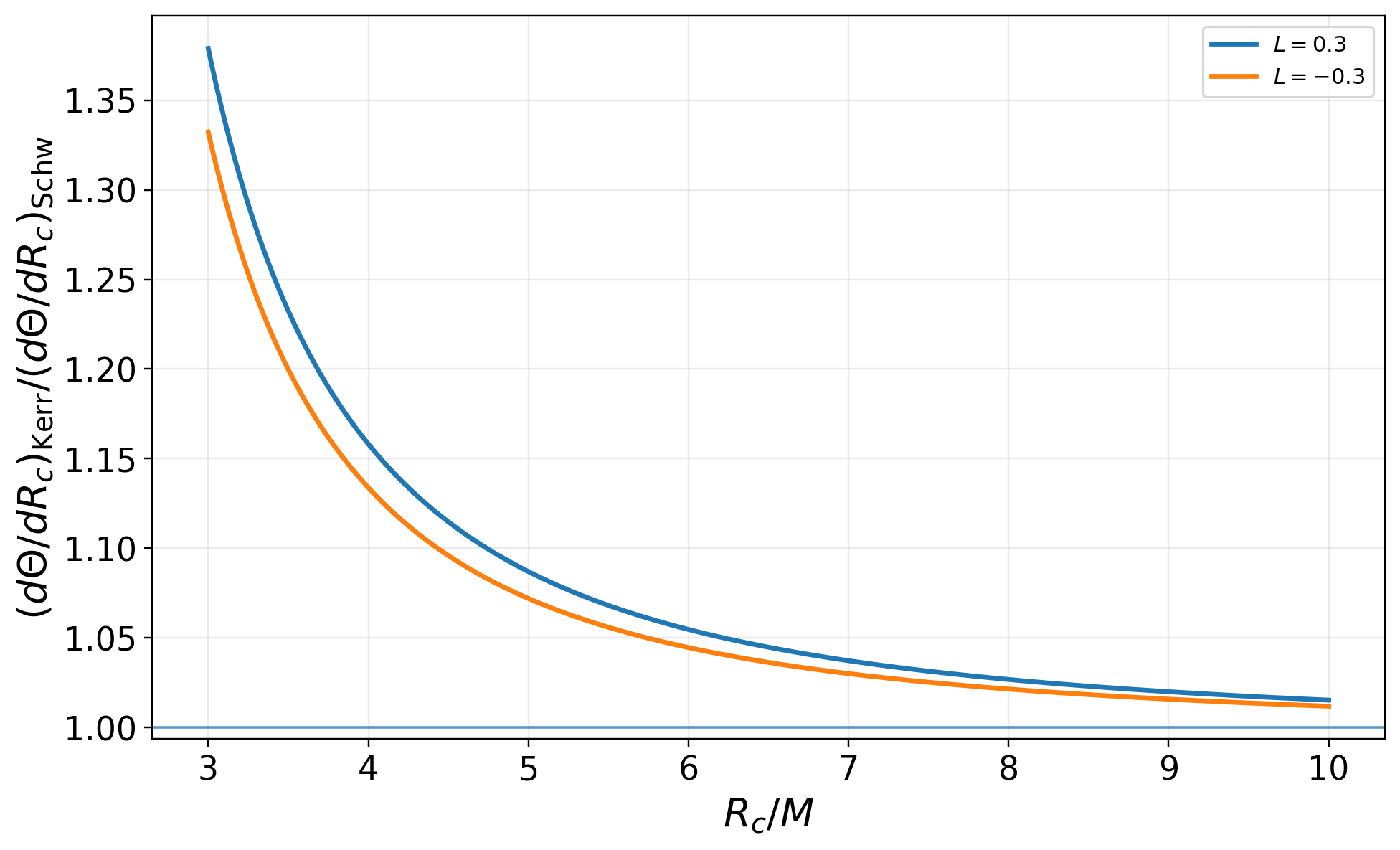}
        \caption{$\frac{d\Theta}{dR_c}$ ratio}
        \label{fig:dtheta_E095}
    \end{subfigure}
    \\
    \begin{subfigure}[t]{0.65\textwidth}
        \centering
        \includegraphics[width=\textwidth]{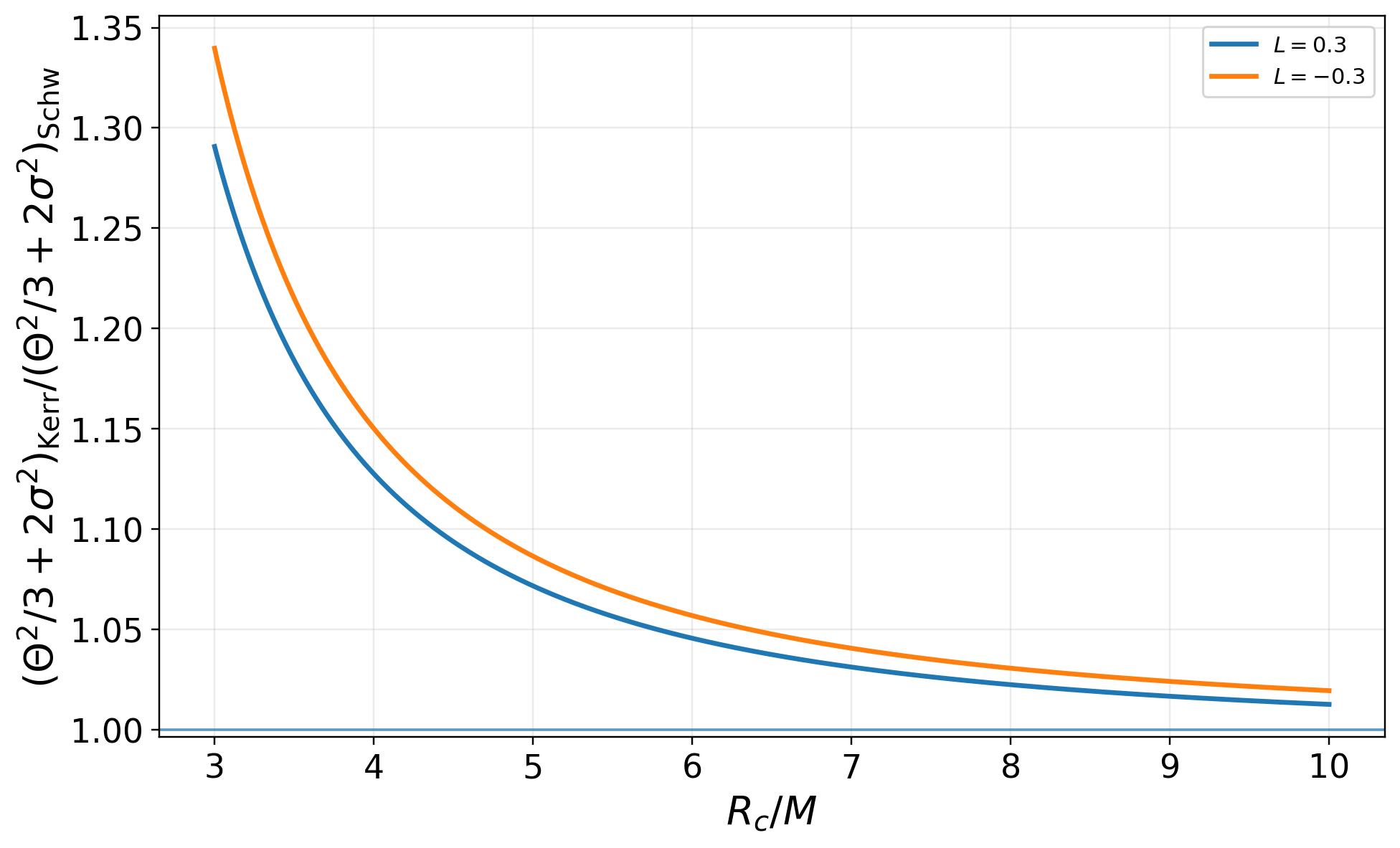}
        \caption{$\frac{\Theta^2}{3}+2\sigma ^2$ ratio}
        \label{fig:dsigma2_E095}
    \end{subfigure}
    \\
    \begin{subfigure}[t]{0.65\textwidth}
        \centering
        \includegraphics[width=\textwidth]{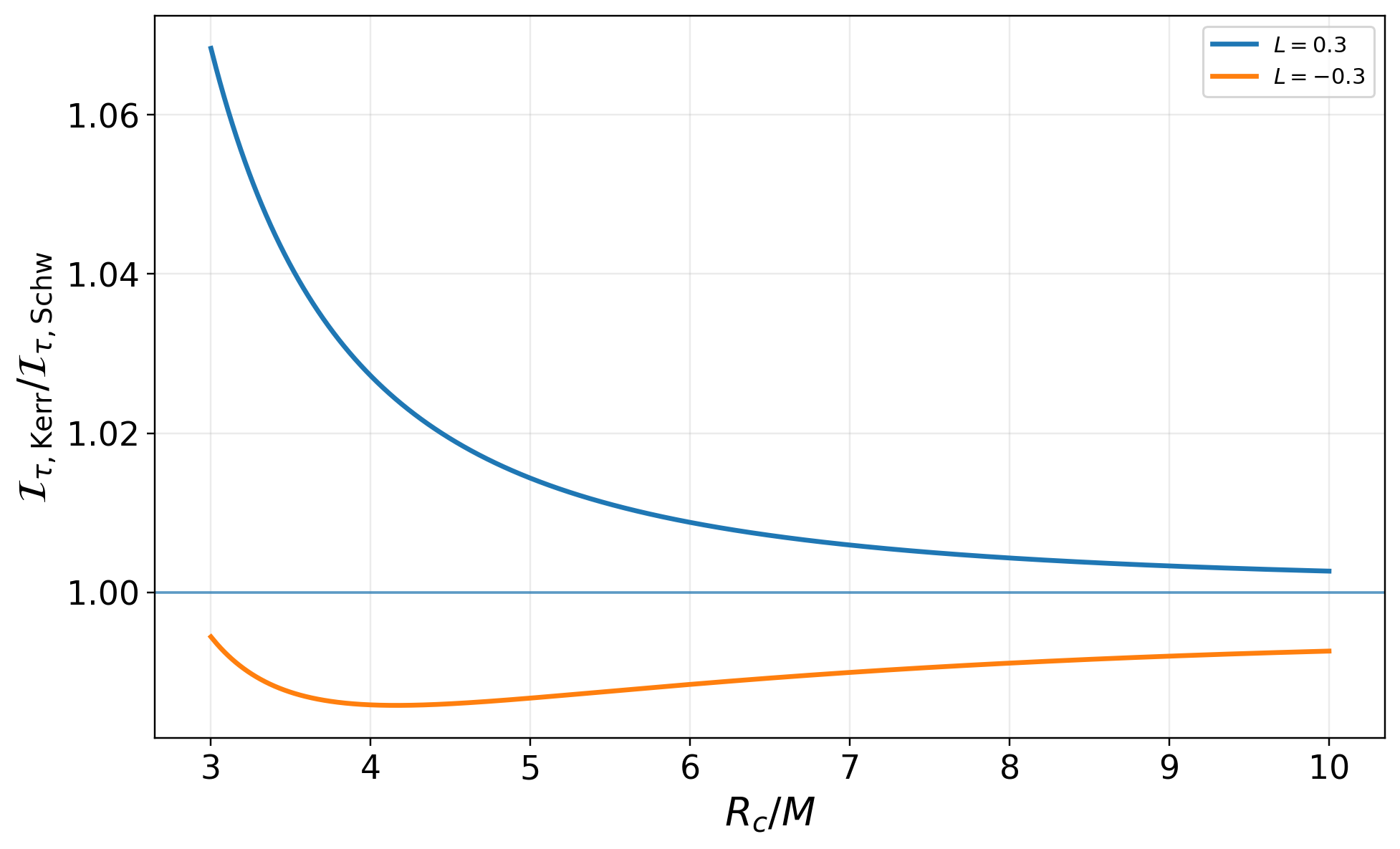}
        \caption{$I_\tau$ ratio}
        \label{fig:dthetadot_E095}
    \end{subfigure}

    \caption{
Covariant Raychaudhuri quantities for equatorial timelike congruences with $E=0.95$, $|L|=0.3$, and $a=0.9$, comparing Kerr and Schwarzschild spacetimes at equal circumferential radius. Panel (a) shows the ratio of the expansion-gradient term $(d\Theta/dR_c)_{\rm Kerr}/(d\Theta/dR_c)_{\rm Schw}$, panel (b) shows the ratio of the focusing contribution $(\Theta^2/3+2\sigma^2)_{\rm Kerr}/(\Theta^2/3+2\sigma^2)_{\rm Schw}$, and panel (c) shows the ratio of the Raychaudhuri time integrand $\mathcal I_{\tau,\rm Kerr}/\mathcal I_{\tau,\rm Schw}$.}
    \label{fig:Raychaudhury_E095}
\end{figure*}

\begin{figure*}
    \centering
    \begin{subfigure}[t]{0.65\textwidth}
        \centering
        \includegraphics[width=\textwidth]{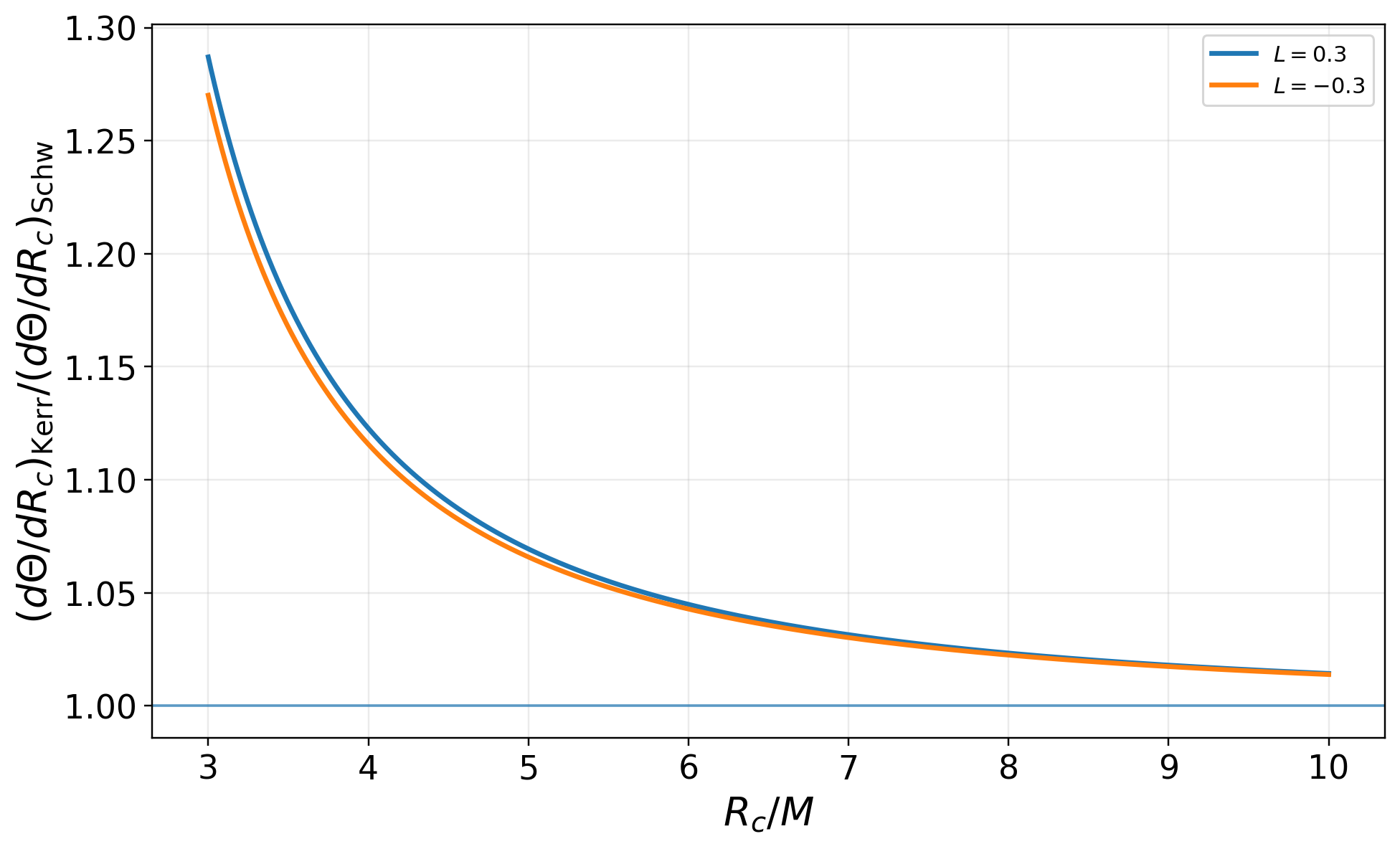}
        \caption{$\frac{d\Theta}{dR_c}$ ratio}
        \label{fig:dtheta_E095}
    \end{subfigure}
    \\
    \begin{subfigure}[t]{0.65\textwidth}
        \centering
        \includegraphics[width=\textwidth]{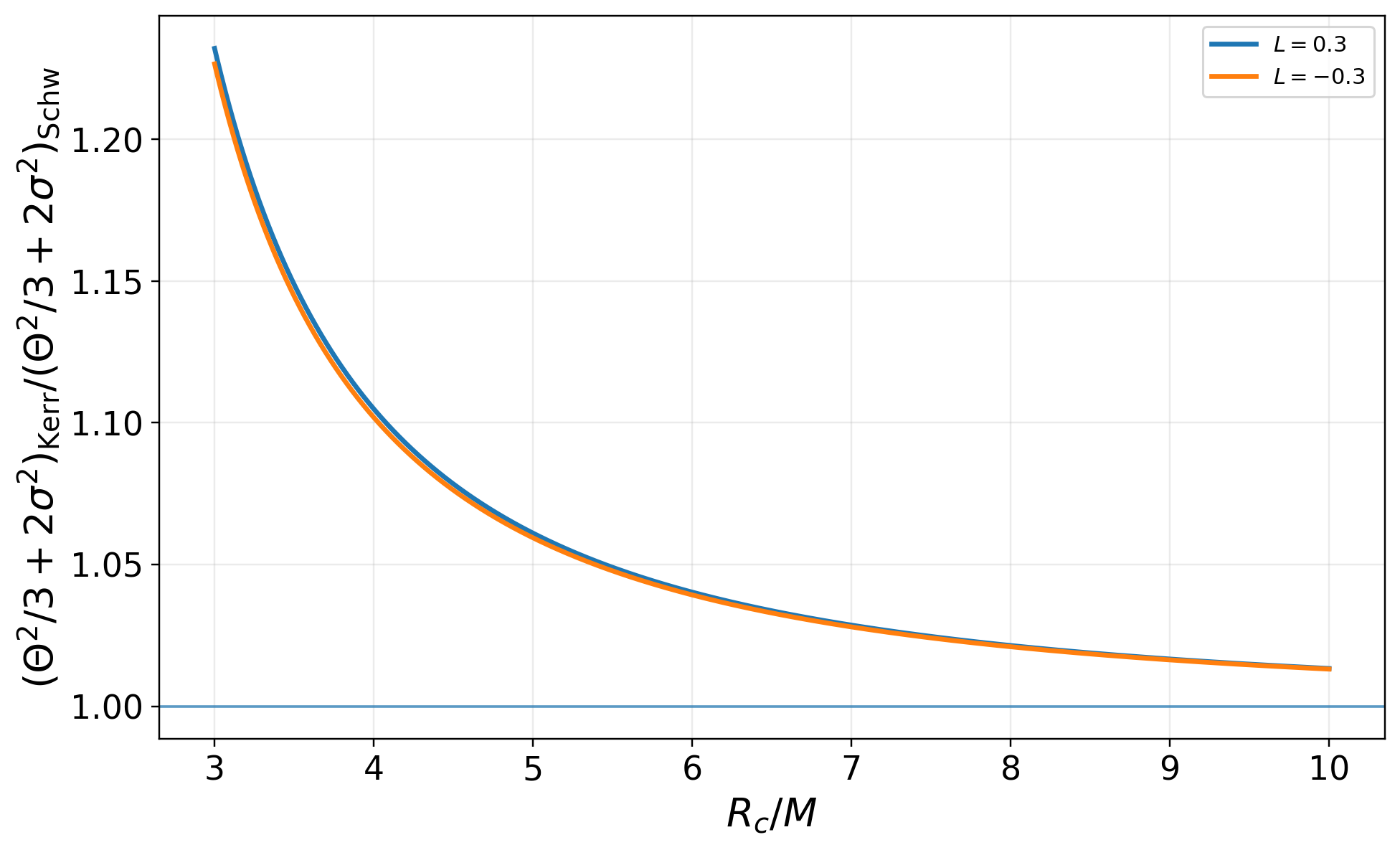}
        \caption{$\frac{\Theta^2}{3}+2\sigma ^2$ ratio}
        \label{fig:dsigma2_E095}
    \end{subfigure}
    \\
    \begin{subfigure}[t]{0.65\textwidth}
        \centering
        \includegraphics[width=\textwidth]{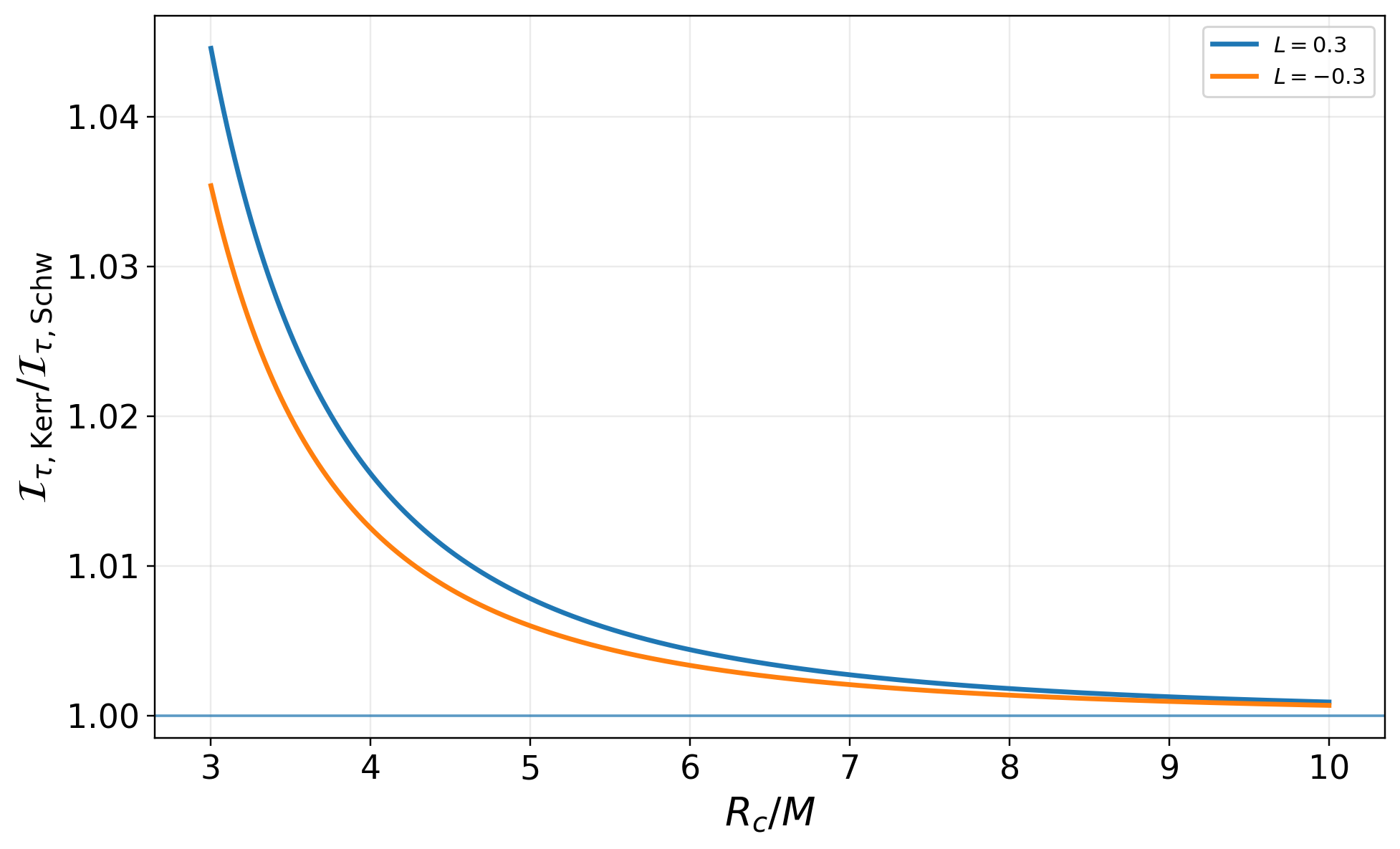}
        \caption{$I_\tau$ ratio}
        \label{fig:dthetadot_E095}
    \end{subfigure}

    \caption{
Same as Fig.~\ref{fig:Raychaudhury_E095}, but for the highly relativistic regime $E=5$.
}
    \label{fig:Raychaudhury_E5}
\end{figure*}

In Figs.~\ref{fig:Raychaudhury_E095} and \ref{fig:Raychaudhury_E5} we illustrate how Kerr rotation modifies the covariant quantities entering the Raychaudhuri time integrand for timelike geodesic congruences with fixed $|L|=0.3$, considering moderately relativistic ($E=0.95$) and highly relativistic ($E=5$) regimes. In both cases, we adopt a high spin parameter $a=0.9$ in order to clearly expose rotational effects.

Panels (a) show the ratio of the expansion-gradient contribution,
\[
(d\Theta/dR_c)_{\rm Kerr}/(d\Theta/dR_c)_{\rm Schw},
\]
while panels (b) show the ratio of the nonlinear focusing term,
\[
(\Theta^2/3+2\sigma^2)_{\rm Kerr}/(\Theta^2/3+2\sigma^2)_{\rm Schw}.
\]
Panels (c) display the resulting ratio of the Raychaudhuri time integrand,
\[
\mathcal I_{\tau,\rm Kerr}/\mathcal I_{\tau,\rm Schw}.
\]

At moderate energies ($E=0.95$), Kerr rotation enhances both the radial evolution of the expansion and the nonlinear focusing contribution relative to Schwarzschild. However, the relative enhancement differs between prograde and retrograde congruences. For prograde trajectories ($L>0$), the increase in the expansion-gradient term dominates over the increase in the focusing contribution, leading to
\[
\mathcal I_{\tau,\rm Kerr}
>
\mathcal I_{\tau,\rm Schw},
\]
and therefore to longer integrated proper infall times. By contrast, for retrograde trajectories ($L<0$), the focusing contribution is comparatively more strongly enhanced, producing
\[
\mathcal I_{\tau,\rm Kerr}
<
\mathcal I_{\tau,\rm Schw},
\]
and consequently shorter infall times relative to Schwarzschild.

At higher energies ($E=5$), the distinction between prograde and retrograde configurations becomes less pronounced. Although Kerr rotation still enhances both the expansion-gradient and focusing terms, the relative balance changes in such a way that the Raychaudhuri integrand becomes larger than its Schwarzschild counterpart for both orbital orientations. This explains why the integrated proper infall times become longer in Kerr spacetime for both prograde and retrograde families at sufficiently large energies.

These results provide a covariant interpretation of the Kerr--Schwarzschild differences in integrated proper infall times. Rather than being controlled by expansion or shear separately, the infall dynamics are governed by the competition between the radial evolution of the expansion and the nonlinear focusing contribution appearing in the Raychaudhuri equation.

\section{Discussion and conclusions}

In this work we have investigated integrated proper infall times in the Schwarzschild and Kerr spacetimes using a covariant $1+3$ framework based on timelike geodesic congruences. Our analysis was motivated by the fact that comparisons between distinct spacetime geometries require specifying which geometric structures are held fixed, since no canonical point-by-point identification exists between different solutions of Einstein's equations.

To address this issue, we compared Schwarzschild and Kerr trajectories connecting surfaces of equal circumferential radius in the equatorial plane. Within this geometrically consistent prescription, we computed integrated proper infall times for different values of the particle energy $E$, angular momentum $L$, and Kerr spin parameter $a$. We showed that black-hole rotation can either increase or decrease the integrated proper infall time relative to Schwarzschild, depending on the orbital configuration and energy regime considered. Those results illustrate that the comparison procedure itself forms part of the physical interpretation when relating distinct relativistic spacetimes.

A central result of this work is that these differences admit a natural covariant interpretation in terms of the Raychaudhuri dynamics of timelike geodesic congruences. Rather than being controlled by expansion or shear separately, the integrated infall times are governed by the competition between the radial evolution of the expansion and the nonlinear focusing contribution associated with expansion and shear. This balance is encoded in the local Raychaudhuri time integrand,
\[
\mathcal I_\tau(R_c)
\equiv
\frac{d\Theta/dR_c}
{\Theta^2/3+2\sigma^2},
\]
whose cumulative contribution along the trajectory determines the final proper infall time.

At moderate energies, Kerr rotation modifies the relative balance between these two effects differently for prograde and retrograde trajectories, leading to either longer or shorter infall times relative to Schwarzschild. At sufficiently high energies, the distinction between orbital orientations becomes less pronounced, and the Kerr contributions tend to increase the Raychaudhuri integrand for both families.

More broadly, our results illustrate that the comparison between Kerr and Schwarzschild dynamics is intrinsically sensitive to the geometric prescription used to identify trajectories across the two spacetimes. Different choices of comparison surfaces may lead to qualitatively different conclusions regarding the integrated dynamics. In this sense, the present work emphasizes the importance of carefully defining geometrically meaningful observables when comparing distinct relativistic spacetimes.

Although the present analysis was restricted to equatorial test-particle geodesics, the framework developed here suggests a possible route toward understanding how spacetime rotation modifies relativistic infall dynamics from a fully covariant perspective. Extensions to more general congruences, non-equatorial motion, and hydrodynamical relativistic flows remain interesting directions for future work.

\section*{Acknowledgements}

The author thanks Christos G. Tsagas for insightful discussions and valuable comments that helped clarify the covariant interpretation of the results presented in this work.

\bibliographystyle{apsrev4-2}
\bibliography{mybib}

\end{document}